# Towards optimal performance and in-depth understanding of spinel Li$_4$Ti$_5$O$_{12}$ electrodes through phase field modeling

Alexandros Vasileiadis[†], Niek J.J. de Klerk[†], Raymond B. Smith[*], Swapna Ganapathy[†], Peter Paul R. M. L. Harks[¥], Martin Z. Bazant[*£], Marnix Wagemaker[†§]

[†] Storage of Electrochemical Energy (SEE), Department of Radiation Science and Technology, Faculty of Applied Sciences, Delft University of Technology, Mekelweg 15, 2629 JB Delft, The Netherlands.

[*] Department of Chemical Engineering, Massachusetts Institute of Technology, Cambridge, Massachusetts 02139, USA.

[¥] Materials for Energy Conversion and Storage (MECS), Faculty of Applied Sciences, Delft University of Technology, P.O. Box 5045, 2600 GA Delft, The Netherlands.

[£] Department of Mathematics, Massachusetts Institute of Technology, Cambridge, Massachusetts 02139, USA.

[§] e-mail: m.wagemaker@tudelft.nl

**ABSTRACT:** Computational modeling is vital for the fundamental understanding of processes in Li-ion batteries. However, capturing nanoscopic to mesoscopic phase thermodynamics and kinetics in the solid electrode particles embedded in realistic electrode morphologies is challenging. In particular for electrode materials displaying a first order phase transition, such as LiFePO$_4$, graphite and spinel Li$_4$Ti$_5$O$_{12}$, predicting the macroscopic electrochemical behavior requires an accurate physical model. Herein, we present a thermodynamic phase field model for Li-ion insertion in spinel Li$_4$Ti$_5$O$_{12}$ which captures the performance limitations presented in literature as a function of all relevant electrode parameters. The phase stability in the model is based on ab-initio DFT calculations and the Li-ion diffusion parameters on nanoscopic NMR measurements of Li-ion mobility, resulting in a parameter free model. The direct comparison with prepared electrodes shows good agreement over three orders of magnitude in the discharge current. Overpotentials associated with the various charge transport processes, as well as the active particle fraction relevant for local hotspots in batteries, are analyzed. It is demonstrated which process limits the electrode performance under a variety of realistic conditions, providing comprehensive understanding of the nanoscopic to microscopic properties. These results provide concrete directions towards the design of optimally performing Li$_4$Ti$_5$O$_{12}$ electrodes.

1. Introduction

Li-ion batteries are widely implemented in mobile and static applications that require high standards of energy and power density, as well as a prolonged and stable cycle life.[1] The most commonly used anode is graphite, a material which usually fails to meet the high rate, safety and cycle-stability requirements for next generation applications.[2] As a result, negative electrode candidates with superior electrochemical characteristics are required. One of the most prominent and well-studied negative electrode materials is the spinel $Li_4Ti_5O_{12}$ (LTO). LTO can be cycled reversibly up to its theoretical capacity of 175 mAh/g[3] and has a stable voltage plateau at 1.55 V vs. $Li/Li^+$[4]. Despite the relatively large potential for a negative electrode and small capacity, compromising the energy density, it allows safe operation inside the stability window of most organic electrolytes. Consequently, dendrite and solid electrolyte interface (SEI) formation is minimized, enabling LTO to meet superior safety standards and performance.[2b, c, 5] Besides the reduction of undesirable side reactions, the exceptional cycle life of LTO is also the consequence of its "zero strain" property.[6] During lithiation the lattice parameters remain almost unchanged, resulting in a negligible volume change of 0.2% [7] minimizing mechanical instabilities induced by lattice strain and volume changes during cycling. In combination with the low costs for preparation[8], the excellent safety and performance characteristics of the LTO material make it an ideal candidate for sustainable energy storage and transport applications.[2c, d]

Although LTO is already applied in Li-ion batteries for modern commercial applications[2b, 9], a number of crucial challenges remain. These include avoiding gas evolution[10] (CO, $H_2$ and $CO_2$) upon cycling, improving the battery energy density and improving the power performance.[11] In an effort to achieve enhanced electrode performance a huge variety of experimental optimization methods has been investigated[2b, d, 12a, b], including doping, surface coating, nanostructuring, and controlling the morphology of the material.

The crystal lattice of the spinel $Li_4Ti_5O_{12}$ is described by the cubic Fd-3m space group (no. 227).[4b] In $Li_4Ti_5O_{12}$ the Li ions reside in the tetrahedral 8a sites, and partially (1/6) in the electrochemically inactive octahedral 16d sites. Upon lithiation the 8a site becomes unstable, and the inserted Li-ions and Li-ions at the 8a sites will occupy the 16c octahedral sites.[13] This topotactic transition causes the structure to undergo a phase transition from a spinel to a rock-salt lattice, reaching a final composition of $Li_7Ti_5O_{12}$. This process is accompanied by a redox reaction reducing $Ti^{+4}$ to $Ti^{+3}$. The material is believed to transform via a core-shell, two-phase separation mechanism between the two spinel endmember phases $Li_4Ti_5O_{12}$ and $Li_7Ti_5O_{12}$.[2d, 3a, 7a, 14a, b] This description is consistent with the exceptionally flat voltage profile which suggests sharp solubility limits ($\delta$ = 0.09, $Li_{4+\delta}Ti_5O_{12}$, $Li_{7-\delta}Ti_5O_{12}$).[15] A detailed recent density functional theory (DFT) investigation revealed that the phase interface is stabilized by adjacent Li-ion occupation in 8a and 16c sites, the mixing of which creates the conditions for fast Li-ion diffusion and excellent electronic conductivity.[16] The phase transition in the material nevertheless remains a first order transition, since the 16c and 8a regions remain separated, albeit on a sub-nanometer length scale[16], consistent with electron microscopy observations which reveal phase separation with a sharp phase interface.[17]

Accurate description of the electrochemistry of Li-ion electrodes requires detailed consideration of the Li-ion kinetics and the thermodynamic properties. Electrochemical phase-field modelling[18] has been shown to be successful in capturing the phase transformation behavior in Li-ion battery materials. In the prototypical case of $LiFePO_4$ (LFP)[18b, 19], the phase field method predicted intercalation waves along the active crystal facet[20], successfully explained the smaller miscibility and spinodal gaps in nanoparticles[21] observed experimentally[22], and predicted the suppression of phase separation during lithium insertion above a critical current[23], which was experimentally observed a few years later[24]. The addition of coherency strain in the Gibbs free energy description enabled the prediction of striped morphologies and phase boundary orientations in LFP nano-particles[21c], which were experimentally validated by microbeam diffraction[25], while enhancing the predicted rate-dependent suppression of phase separation. Further addition of composition-dependent surface energy led to a theory of size-dependent nucleation in nanoparticles[26] and explained the slightly tilted plateau in the open circuit voltage when integrated in a porous electrode model[18c]. Furthermore, the same general framework of multiphase porous electrode theory[18b, c, d] (MPET) predicted the transition from particle-by-particle mosaic to concurrent intercalation in LFP observed by *in situ* x-ray and electron imaging[27]. Properly accounting for experimentally observed mosaic instabilities in porous graphite anodes[28] is an essential first step toward the prediction of capacity fade and aging due to side reactions, such as SEI growth and lithium plating, which depend on electrochemical heterogeneities.

Recently, the concentration evolution of individual LFP nanoparticles has been observed *in operando* during cycling under realistic conditions by scanning tunneling x-ray microscopy (STXM), which paves the way for unprecedented validation and application of phase-field models at the nanoscale[29]. The local exchange current density versus concentration was extracted from a massive dataset of STXM movies, and its behavior in the spinodal region, coupled with direct observations of phase separation, confirmed the theory that phase separation is suppressed by auto-inhibition during insertion and enhanced by auto-catalysis during extraction[30]. With advances in 3D simulations of LFP nanoparticles[20b, 21b], a comprehensive picture of the lithium intercalation is emerging. The same approach is beginning to be applied to anode materials, such as graphite[19, 31a, b] and $TiO_2$ anatase[32], which exhibit multiple phase transitions.

Relatively few models have been developed for LTO, most having a specific system focus and limited range of parameter coverage.[33] Kashkooli et. al.[33a] matched a model with experiments utilizing LTO monodisperse nanoparticles. He focused on the effect of nanosizing for optimal electrode performance covering rates up to 5C. Stewart et. al.[33c] focused on optimizing a LFP-LTO battery system and an asymmetric hybrid super-capacitor consisting of LTO and activated carbon. The aim was to investigate if the systems meets the requirements for hybrid electric vehicles applications. Christensen et. al[33b] also focused on full-cells consisting of $LiMn_2O_4$ and LTO, investigating the effect of porosity and thickness as well as the particle size. All previous LTO models are based the pioneering work of Newman and collaborators, which focuses on the description of Li-ion diffusion, but does not include a thermodynamic description of phase-separating electrode materials.

In the present work we develop a phase-field model for LTO based on nonequilibrium electrochemical thermodynamics, where all necessary physical and chemical parameters are based on available experimental and ab-initio results. The developed model is able to reproduce presented and published experimental results and is able to capture and explain the fine balance between the various possible limiting factors in the battery, including Li-ion diffusion, Li-ion transport and electronic transport. Thereby our work validates the physical principles of the phase field method, and the importance of including a complete thermodynamic description in battery models. This allows us to provide concrete guidelines towards optimized electrode performance, guiding the design of optimal electrodes for Li-ion batteries. In addition, in-depth understanding of the lithiation mechanism is achieved by following the active particle fraction during discharge.

## 2. Methods

### 2.1. Phase Field Model for LTO

#### 2.1.1. Model Overview

Our phase-field model of LTO is based on nonequilibrium electrochemical thermodynamics[18a], and simulations are carried out for porous LTO electrodes using the MPET open-source software package[18c] that implements Multiphase Porous Electrode Theory[18b, d]. As in classical porous electrode theory, the battery is conceptually divided into a collection of finite electrolyte volumes, which include the separator and electrode. All of the electrolyte volumes are connected in series, and thus describe the depth information of the simulated battery. The LTO active material is modeled as a collection of solid particles that are in contact with the electrolyte ion reservoir, having a parallel connection within an electrolyte volume. The porosity ($\varepsilon$) represents the volume fraction taken up by electrolyte in a single volume. The thermodynamics inside the particles are described by discretization into solid grid volumes along the intercalation coordinate ($x$-axis). An overview of the model is shown in **Figure 1**.

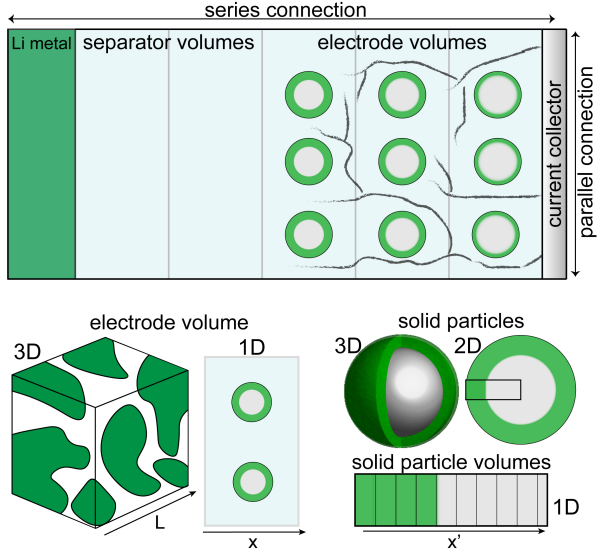

**Figure 1:** Schematic model overview: Here the battery (half-cell) is divided into two separator and three electrode volumes, solid particles are submerged into the electrolyte bath and act as Li-ion sinks/sources.

The system is fully characterized by the determination of several field variables at each time step throughout the (de)lithiation process, namely the electrolyte potential $\varphi_{lyte}$, the electrolyte concentration $c_{lyte}$, the lithium concentration in the solid $c_s$, the potential in the solid $\phi_s$, and the diffusional chemical potential for intercalated lithium $\mu_{Li,LTO}$. The latter represents the free energy change per interstitial position upon lithium intercalation[18a, 23] and is related to the driving force of the following reaction:

$$Li_4Ti_5O_{12} + e^- + xLi^+ \leftrightarrows Li_{4+x}Ti_5O_{12} \quad (1)$$

To capture the thermodynamic behavior of the electrode material, the regular solution model[18a] is implemented based on its ability to probe solid solution vs. phase separating systems. The Gibbs free energy per site, $g(\tilde{c}_s)$, is given by the following equation:

$$g(\tilde{c}_s) = k_B T(\tilde{c}_s \ln(\tilde{c}_s) + (1 - \tilde{c}_s)\ln(1 - \tilde{c}_s)) + \Omega_a \tilde{c}_s(1 - \tilde{c}_s) + \tilde{c}_s \mu_i^\Theta + \frac{1}{2}\frac{\kappa}{c_{max}}|\nabla \tilde{c}_s|^2 \quad (2)$$

where $k_B$ is Boltzmann's constant, $T$ the temperature in Kelvin, $\tilde{c}_s$ the normalized concentration in the solid particles, $\Omega_a$ the enthalpy of mixing, $\mu_i^\Theta$ the equilibrium potential versus Li/Li$^+$ (-1.55eV), $\kappa$ the gradient penalty parameter and $c_{max}$ the maximum concentration of Li. The enthalpy of mixing ($\Omega_a$) quantifies the interactions of intercalated Li-ions within the host material. If $\Omega_a < 2k_B T$ the system favours a solid solution insertion mechanism, while if $\Omega_a > 2k_B T$ the enthalpy term overcomes the competing entropy term leading to phase segregation. The last term is the gradient penalty term that penalizes steep gradients which occur at phase interfaces. The gradient penalty parameter ($\kappa$) can be linked to the width $\lambda_b$ and interfacial tension $\gamma_b$ of a diffuse phase boundary through the following formula[18a]:

$$\kappa \propto \lambda_b^2 c_{max} \Omega_a \propto \frac{\gamma_b^2}{c_{max}\Omega_a} \quad (3)$$

The larger the gradient penalty $\kappa$, the greater the interfacial tension between phases, and, for a given particle size, the less likely the material is to phase separate. By variationally differentiating the Gibbs free energy with respect to the concentration, the diffusional chemical potential ($\mu_{Li,LTO}$) of lithium inserted in LTO can be obtained.

$$\mu_{Li,LTO} = \frac{\partial g}{\partial c_s} - \nabla \cdot \frac{\partial g}{\partial \nabla c_s} = \mu^\Theta + k_B T \ln\left(\frac{\tilde{c}_s}{1-\tilde{c}_s}\right) + \Omega_a(1 - 2\tilde{c}_s) - \frac{\kappa}{c_{max}}\nabla^2 \tilde{c}_s \quad (4)$$

Assuming linear irreversible thermodynamics, the Li-ion flux in the solid particles is proportional to the gradient of the diffusional chemical potential $\nabla \mu_{Li,LTO}$ and can be expressed with a vacancy diffusion mechanism as[18a]:

$$F = -\frac{D_{Li}}{k_B T} c_s (1 - \tilde{c}_s) \nabla \mu_{Li,LTO} \quad (5)$$

where $D_{Li}$ is the tracer diffusivity of isolated Li-ions in the LTO lattice, which is assumed to be a scalar for isotropic diffusion. Since LTO also has negligible coherency strain (which usually introduces strong crystal anisotropy), we further assume spherically symmetric radial diffusion and core-shell phase separation, which is a convenient and reasonable simplification of the single-particle model for LTO, and allows for more efficient simulations.

The charge transfer reaction current ($I$) controls the Li-ion flow between the electrolyte and the solid particles. The generalized Butler-Volmer equation[18a, b], is implemented to describe the Li-flux at the electrolyte/solid boundary resulting from Faradaic intercalation reactions:

$$I = k_0 \frac{a_{Li,lyte}^{1-\alpha} a_{Li,s}^{\alpha}}{\gamma_{TS}} \left[ exp\left(-\frac{\alpha e \eta_{eff}}{k_B T}\right) - exp\left(\frac{(1-\alpha) e \eta_{eff}}{k_B T}\right) \right] \quad (6)$$

where $k_0$ is the rate constant per intercalation site, and $\alpha$ is the symmetry factor of the activation barrier along the reaction coordinate. The exchange current depends upon the activity of lithium, both in the electrolyte ($a_{Li,lyte}$) and solid ($a_{Li,s}$). The choice of transition state activity coefficient[18a], $\gamma_{TS} = (1 - \tilde{c}_s)^{-1}$, which assumes one excluded site for the transition state during lithium insertion, leads to an asymmetric exchange current versus concentration, peaking at low concentrations, which has recently been validated by local measurements of the intercalation rate in LFP nanoparticles[29]. As a result, the insertion reaction is auto-inhibitory (and the extraction reaction is auto-catalytic) across the spinodal region, which leads to suppression of phase separation during insertion (and enhancement during extraction), as predicted theoretically[23, 30] and later confirmed by *in operando* imaging of nanoparticle concentration profiles[29]. The effective overpotential $\eta_{eff}$ includes film resistance ($R_{film}$) in series with the Faradaic reaction[18d],

$$\eta_{eff} = \eta + I R_{film} \quad (7)$$

which helps to accurately fit experimental data with MPET[28], as well as with classical porous electrode models. This common approximation yields a curved Tafel plot, which could result from an actual series reaction resistance, such as a thin film or Stern layer hindering electron or ion transport to the reaction site. Alternatively, the curved Tafel plot may signify quantum mechanical deviations from Butler-Volmer kinetics[18d], as predicted by the Marcus theory of electron transfer and consistent with experiments on LFP[34].

The overpotential $\eta$ is calculated through its local thermodynamic definition:

$$\eta \equiv \Delta \phi - \Delta \phi^{eq} \quad (8)$$

with $\Delta \phi = \phi_s - \phi_{lyte}$ being the potential difference between the solid electron-conducting phase and the electrolyte at the electron-transfer reaction site. The equilibrium potential $\Delta \phi^{eq}$ is determined through the Nernst equation:

$$\Delta \phi^{eq} \approx -\frac{\mu_{Li,LTO}}{e} + \frac{k_B T}{e} \ln \frac{c_{lyte}}{c_0} \quad (9)$$

where we neglect Frumkin corrections associated with charged double layers[35], by setting $\varphi_{lyte}$ and $c_{lyte}$ to their nearly constant local values in the bulk electrolyte, which are assumed to vary only over the macroscopic electrode length scale. We also neglect concentration-dependent variations of the liquid-state Li-ion activity coefficient in the Nernst equation, since our focus is on the larger variations of Li-ion activity in the solid captured by the phase-field model. The Li-ion transport in the electrolyte is described using the Stefan-Maxwell concentrated electrolyte model.[18d] A linear Ohmic-like function was implemented to describe electronic wiring in the electrode[18d]:

$$j_{electronic} = -\sigma \nabla \phi_s \quad (10)$$

where $\sigma$ is the electronic conductivity.

The coupled partial differential equations (PDE's) are solved simultaneously throughout the system using a finite volume discretization method. In the current simulations the applied current was fixed, which allows the determination of the voltage and other related variables (e.g. lithium diffusional potential) of the simulated battery during Li (de)intercalation. For simplicity the LTO particles are described as one-dimensional, as shown in **Figure 1**, which appears appropriate based on the isotropic nature of the Li-ion diffusion in LTO. During the simulations the effects of coherency strain, which are important for most intercalation materials such as LFP[21b, c, 26], were neglected motivated by the special "zero-strain" property of LTO.[4a, b, 7a]

### 2.1.2. Parametrizing for LTO

To capture the nature of the phase transition in electrode materials requires the enthalpy of mixing ($\Omega_a$), which in most thermodynamic models is obtained by fitting the equilibrium voltage profile. Based on the equilibrium voltage profile of LTO[4b, 36] this results in: $\Omega_a = 1.52 \cdot 10^{-20} \, J/Li$, a value corresponding to a phase separation between x=0.09 and x=2.91 in $Li_{4+x}Ti_5O_{12}$.[15] Aiming at reducing the amount of fitted parameters, a more direct way to obtain the thermodynamic properties is to determine the enthalpy of mixing based on extensive DFT calculations recently published[16] The convex hull determined for lithium insertion in LTO is presented in **Figure 2a**.

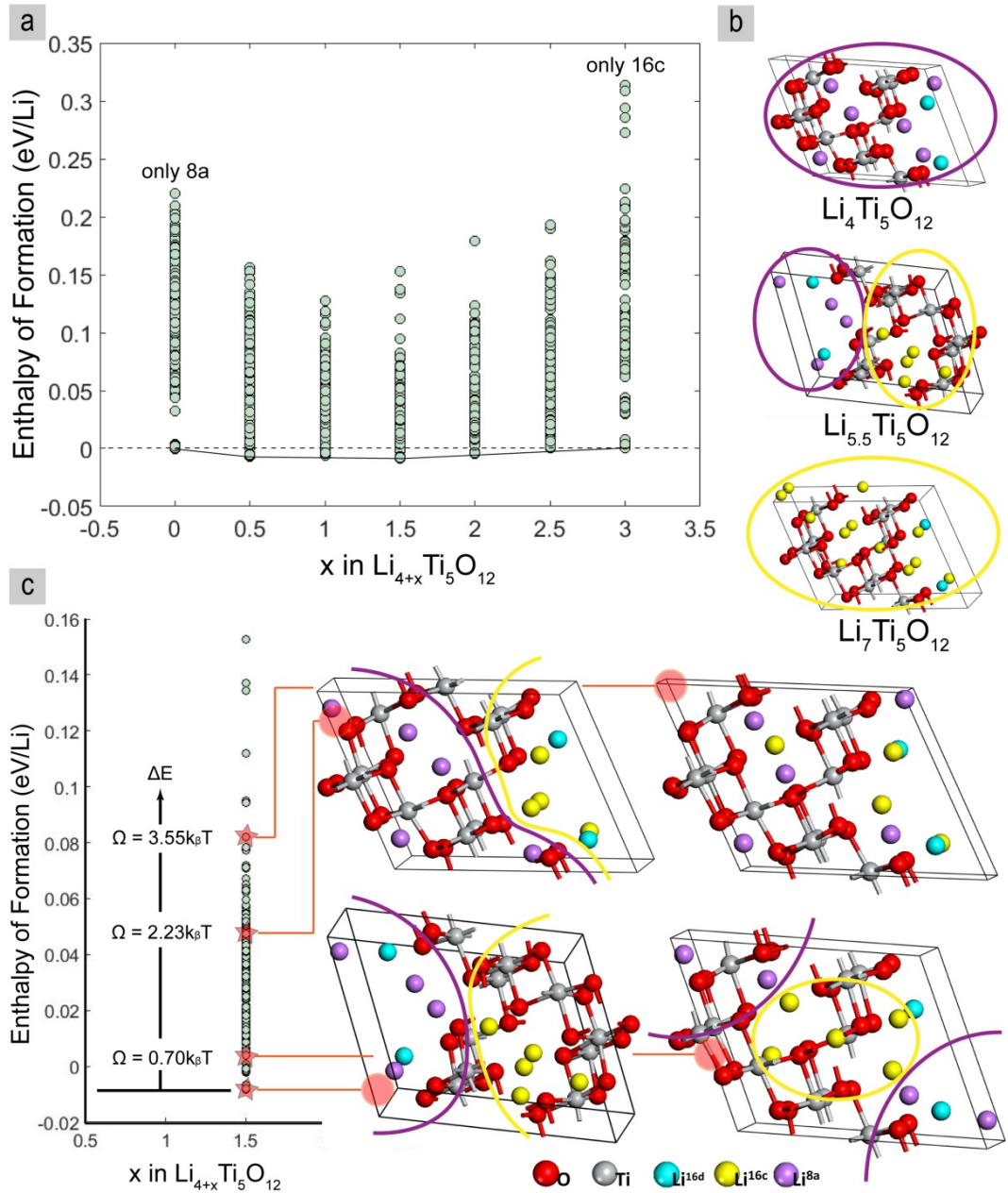

**Figure 2: a)** Enthalpies of formation for Li-ion insertion in LTO and the resulting convex hull, the line connecting the lowest energy configurations referenced to the $Li_4Ti_5O_{12}$ (all electrochemically active Li-ion on the 8a positions) and $Li_7Ti_5O_{12}$ (all electrochemically active Li-ion on the 16a positions) end-member compositions. **b)** Lowest energy configuration in the end-member and intermediate concentrations. **c)** Enthalpies of formation for the $Li_{5.5}Ti_5O_{12}$ phase. The enthalpy of mixing is defined as the enthalpy difference (ΔE) between the lowest energy configuration and the lowest one without a clear phase-interface.

In **Figure 2a** it is shown that the formation enthalpy of several configurations at intermediate concentrations are positioned on the convex hull, suggesting solid solution behavior for $Li_{4<x<7}Ti_5O_{12}$. However, upon examining lowest energy configurations, lithium clustering into 8a and 16c regions is observed within single unit cells, representing a sub-nanometer phase separation[16]. An example of this is shown in **Figure 2b,** in which the lowest energy configuration of the $Li_{5.5}Ti_5O_{12}$ phase consists of phase separated regions of the end-member phases. The enthalpy difference between the lowest enthalpy configuration (displaying phase separation) and the configurations in which the 16c and 8a Li-ion occupation is mixed (displaying solid solution) provides a quantification of the enthalpy of mixing. This is demonstrated in **Figure 2c** for $Li_{5.5}Ti_5O_{12}$, where the Li-positions are shown for several configurations with different formation enthalpies.

The difference in formation energy between the most stable configuration (un-mixed) and the completely mixed configuration for $Li_{5.5}Ti_5O_{12}$ results in $\Omega_a = 1.43 \cdot 10^{-20} \, J/Li$. This results compares very well with fitting result of the enthalpy of mixing towards the equilibrium voltage profile, motivating the use of this value for the phase field model. The interface width ($\lambda$) in LTO has been determined by experimental[17, 37] and computational[16, 38] studies, both indicating a narrow interface width of a few Angstroms wide (2.5 to 7 Å). Using equation 3 this results in a gradient energy penalty of $\kappa = 1 \cdot 10^{-10} \, J/m$, reflecting that the energy penalty for creating an interface between the two endmember phases is relatively small, for comparison, approximately 5 times smaller compared to the two coexisting phases in $LiFePO_4$.

Li-ion diffusion in the spinel LTO lattice has been extensively studied with nuclear magnetic resonance (NMR) spectroscopy[15, 39a, b, c], DFT [16, 40a, b] and electrochemical methods[3b, 41] reporting a range of chemical diffusivities from $10^{-8}$ to $10^{-12}$ cm²/s. NMR[15, 39a, b, c] and DFT[16, 40a, b] studies consistently reveal excellent diffusivities for intermediate Li-ion concentrations, $Li_{4<x<7}Ti_5O_{12}$, and poor diffusivities for the end-member phases $Li_4Ti_5O_{12}$ and $Li_7Ti_5O_{12}$. It is suggested[15, 39a] that the intimate mixing of the 8a and 16c domains is responsible for the dramatic Li-ion diffusion enhancement at intermediate concentrations. This was also predicted by recent DFT molecular dynamic simulations, demonstrating that Li-ions are very mobile at the phase interfaces, the abundance of which is enhanced by the interface stabilization through Li-ion on the 16d metal positions.[16] For the current phase field model the (tracer) self-diffusion coefficient was used as determined by NMR spectroscopy amounting to $D_{Li} = 4 \cdot 10^{-12} \, cm^2/s$[15], consistent with values determined by other NMR investigations[20a, b] and electrochemical measurements[4a, 41d, e].

Particle aggregation and agglomeration is a commonly observed for LTO materials when synthesized via conventional techniques[2d, 41e, 42]. Because the diffusion length of the Li-ions is determined by the particle size this has a large effect on the time scale of Li-ion insertion and extraction, and therefore it is essential to accurately represent experimental particle size distributions. A lognormal description was found to adequately capture experimentally observed grain size distributions[43a, b]. Due to the long "tail" of the lognormal function the presence of a few big aggregates is ensured, which, as will be shown, have an important effect on electrode performance. To manage computational time the number of particles in the phase field model was restricted to be between 35 and 200, depending on the particle size, having a size distribution consistent with the log-normal distribution determined from SEM images, see **Supporting Information A**.

The electronically insulating nature of spinel $Li_4Ti_5O_{12}$[36, 44] resulted in considerable experimental effort to improve the electronic conductivity of LTO electrodes[2b, 12b, 45]. However, Song et al. indicated that $Li_4Ti_5O_{12}/Li_7Ti_5O_{12}$ phase interfaces are highly electronically conducting[46] and in line with this carbon free LTO batteries were tested showing excellent rate capability[46, 47]. Conductivity measurements during electrochemical cycling of LTO report a tremendous increase in electronic conductivity of several orders of magnitude at the very first stages (4%) of the lithiation process[47, 48]. Thus it was suggested that either the Li-ion diffusion in the solid phase[48] or Li-ion transport through the electrolyte in the pores of the electrode[47, 49] may limit the rate performance. Previous modeling attempts integrated both extremely small and large values of the electronic conductivity. Christensen et. al,[33b] assumed that matrix conductivity is not rate limiting due to the addition of carbon used an arbitrary value of 100 S/m while Kashkooli et al.[33a] implemented a value of just 0.02 S/m. In the current phase field model an effective electronic conductivity of 10 S/m is implemented, which is a typical value for a 10% conductive additive. Additionally, in order to investigate the influence of the effective electronic conductivity, further calculations with larger and smaller values were performed.

Tortuosity quantifies the effect of the electrode microstructure on the kinetic properties of the electrolyte in the electrode pores. The generalized Bruggeman relation[50], relating the porosity ($\varepsilon$) with the tortuosity ($\tau$) via the Bruggeman exponent (p), was implemented.

$$\tau = \varepsilon^p \quad (11)$$

The Bruggeman exponent was set to -0.5, which corresponds to the ideal case for spherical particles. Additionally, the effect of the tortuosity on the capacity of LTO electrodes was studied.

### 2.1.3. Analysis of the total overpotential

The various contributions to the total overpotential were estimated in order to gain insight on the rate limiting mechanism of LTO electrodes. The contributions were measured by simulating the respective electrode at a certain rate, while switching off one kinetic mechanism (transport, diffusion, electronic, transfer) at a time and measuring the potential gain compared with the voltage obtained from an actual simulation. The kinetic mechanisms were switched off by increasing the related property by several orders of magnitude (for example electronic conductivity was increased from 10 to 1000 S/m). Due to the interconnected character of the kinetic mechanisms adding up the overpotential contributions coming from switching off one mechanism at a time adds up to ~104% of the original voltage line. This is sufficient to ensure that no significant information has been lost and that the rate limiting mechanism has been correctly determined.

## 2.2. Experimental Section

*Electrode characterization*

Non-carbon-coated, commercially available LTO (Süd-Chemie AG), with a primary particle size of approximately 140 nm, was used. The LTO particle size distribution was measured using a scanning electron microscope (SEM, Philips/FEI XL 40 FEG), using an acceleration voltage of 15 kV. The porosity was calculated based on the geometry and weight of the electrodes and the density of the constituents. The electrodes were cut into discs with a fixed diameter and their thickness and weight were carefully determined using a digital indicator and an analytical balance, respectively.

*Electrochemical testing*

The electrodes consisted of 70 w% LTO powder (Süd-chemie), 15 w% carbon Super P (Timcal) and 15 w% polyvinylidene fluoride binder. The electrodes were tested in half-cells with lithium metal as reference/counter electrode and a fiber-glass separator. The separators were soaked with electrolyte (1 M $LiPF_6$ in 1:1 vol% ethylene carbonate/dimethyl carbonate). The cells were assembled inside an Ar filled glovebox with oxygen and water content below 1 ppm. Galvanostatic cycling experiments were performed with a programmable Maccor 4000 series galvanostat. The electrodes were cycled between 1 and 2.5 V vs. Li/Li$^+$ at various C-rates (1C = 175 mA g$^{-1}$).

## 3. Results and Discussion

The default parameters used for the phase field simulations are summarized in **Table 1** unless otherwise indicated. Electrode thickness and porosity will be specified for the respective electrodes in the following sections. Note that all the thermodynamic parameters are obtained from literature. In combination with a known electrode geometry it allows us to run the simulations without fitting any parameters. All simulations were performed at 25 degrees Celsius.

**Table 1:** Default parameters for the LTO phase field model

| **Thermodynamic Parameters** | Value | Based on |
|---|---|---|
| Enthalpy of Mixing ($\Omega_a$) | $1.43 \cdot 10^{-20}\ J/Li$ | DFT *[16] |
| Gradient Penalty ($\kappa$) | $1 \cdot 10^{-10}\ J/m$ | TEM[17], DFT [16, 38] |
| Equilibrium potential ($\mu^\Theta$) | 1.55 V | Electrochemical experiments[4a, b, 15] |
| LTO density ($\varrho_{LTO}$) | $3.5\ g/cm^3$ | |
| **Kinetic Parameters** | | |
| Li-ion Tracer Diffusivity ($D_{Li}$) | $4 \cdot 10^{-16}\ m^2/s$ | NMR[15] |
| Effective Electronic Conductivity ($\sigma$) | $10\ S/m$ | Electrochemical experiments[48] |
| Electrolyte Anion Diffusion ($D_-$) | $2.94 \cdot 10^{-10}\ m^2/s$ | Commercial LiPF6 EC/DMC |
| Electrolyte Cation Diffusion ($D_+$) | $2.2 \cdot 10^{-10}\ m^2/s$ | Commercial LiPF6 EC/DMC |
| Reaction Rate Constant ($k_0$) | $3.6\ A/m^2$ | Electrochemical experiments[51] |
| **Geometry Parameters** | | |
| Particle Morphology | Spherical | SEM images ** |
| Particle Radius ($R$) | $45 - 280\ nm$ | SEM images ** |
| Separator Length ($L_s$) | $200\ \mu m$ | Commercial Whatman GF *** |
| Separator Porosity ($\varepsilon_s$) | 87 % | Commercial Whatman GF *** |
| LTO Volume Loading ($PL_c$) | 57 % | Synthesis process * |
| Bruggeman Exponent ($p$) | -0.5 | Spherical morphology** |

*this work, **Supporting Information, ***effective property when confined in a battery setup[52]

### 3.1. Single LTO particles

Single particle simulations, using the parameters shown in **Table 1**, result in phase separation during lithiation, forming a lithium-rich phase close to the surface and a sharp phase-interface that propagates towards the centre of the particle as shown in **Figure 3a**. The calculated voltage profile shown in **Figure 3a** displays the characteristic voltage plateau for LTO at 1.55 V vs. Li metal for a slow rate discharge (0.1C). The voltage dip observed at the early stages of lithiation is a known attribute of single particle electrochemical behaviour (in the absence of heterogeneous nucleation at surfaces[26]), because the lack of thermodynamic noise fails to push the Gibbs free energy out of the metastable state (by homogeneous nucleation) until the spinodal point is reached.[53] For phase-separating porous electrodes, the near-equilibrium voltage profile will be flatter, with some overshooting of the voltage plateau and oscillations due to many-particle mosaic instabilities[19, 53].

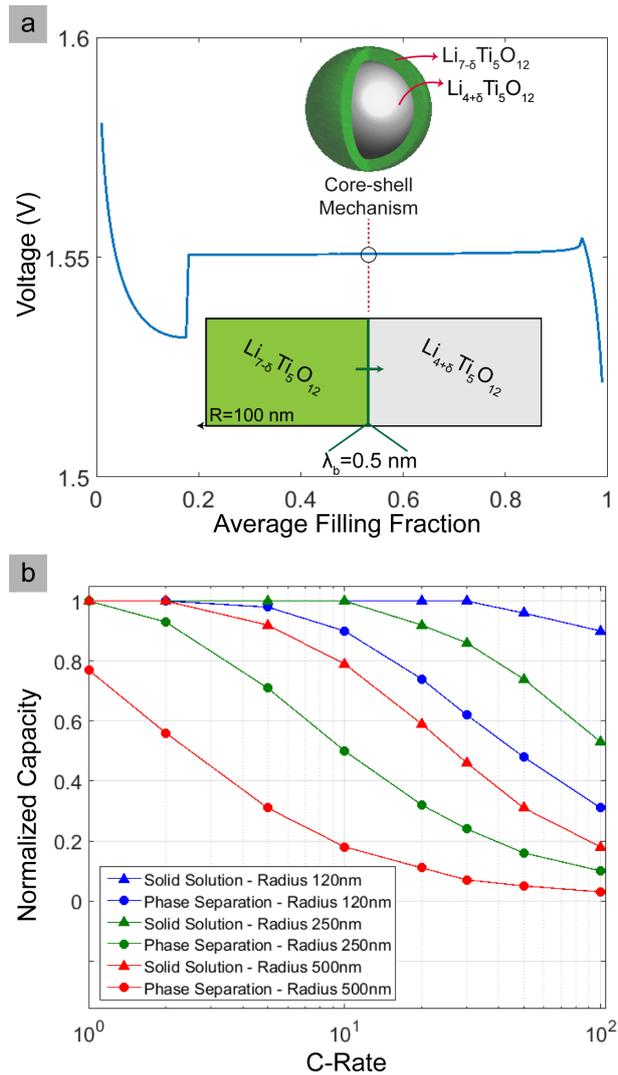

**Figure 3: a)** Galvanostatic discharge voltage profile of a single LTO particle at 0.1C and lithium concentration profile in the solid particle showing a core-shell-like, two-phase separation mechanism with a very thin phase-interface width. **b)** Difference in maximum capacity of single particles between a solid solution description ($\Omega_a = 0.6 \cdot 10^{-20} J/Li$) and a stable phase separation ($\Omega_a = 1.43 \cdot 10^{-20} J/Li$) Note that the x-axis is logarithmic.

A core-shell, two-phase separation description is the generally accepted lithiation mechanism for LTO[2d, 3a, 7a, 14a, b, 54] and is successfully captured by implementing the enthalpy of mixing value from DFT calculations. In order to understand the importance of the phase separation and how it affects the electrochemical performance, phase separation is compared with the solid solution mechanism. In **Figure 3b** the maximum obtainable capacity of LTO single particles is presented for the case of a solid solution description ($\Omega_a = 0.6 \cdot 10^{-20} \, J/Li < 2kT$) vs. a phase separation description ($\Omega_a = 1.43 \cdot 10^{-20} \, J/Li > 2kT$) for three particle sizes at different C-rates. The difference is substantial. A solid solution description leads to high lithiation levels, even for bulk particles as big as 1 $\mu m$, which can be filled to 80% at 10C. On the other hand, implementing a stable phase separation, as indicated by DFT calculations, slows down lithiation significantly, and the 1 $\mu m$ particle is able to reach only 20% capacity at 10C.

The effect of the enthalpy of mixing $\Omega_a$ can be understood as follows. Stable phase separation will lead to the formation of a high concentration phase ($Li_{7-\delta}Ti_5O_{12}$) near the surface of the particle. A high concentration at the surface would increase the reaction resistance by lowering the exchange current (equation 6), and this would lead to larger overpotentials and lower cell voltage, before any diffusion limitation sets in. Eventually, the measured voltage is expected to be lower at higher rates. Furthermore, the existence of a high concentration end-member phase will reduce the Li-ion diffusion (equation 5) due to the lower Li-vacancy concentration. As a result, the more stable the phase separation ($\delta \rightarrow 0$ for $Li_{7-\delta}Ti_5O_{12}$ and $Li_{4+\delta}Ti_5O_{12}$) the easier it is to reach diffusion limitations in the particles. This description is thus able to capture the sluggish Li-ion diffusion within the end-member phases, as is consistently reported by NMR[15, 39a, b, c, 55] and DFT[16, 40a, b] studies. Recently, the fitting of a LTO model without a phase field thermodynamic description to experimental results[33a] required assuming a relatively small diffusion coefficient of $D_{Li} = 2.3 \cdot 10^{-13} \, cm^2/s$, which is about an order of magnitude below NMR experimental results[15, 39b, 55]. Interestingly, it has been argued that the enormous range of diffusivities and exchange currents reported for LFP is also due to improper accounting for phase separation[19]. Similarly, to match the capacity of the phase separating description, the diffusion coefficient in the solid solution phase field model needs to be decreased to an unrealistically small $D_{Li} = 3.8 \cdot 10^{-13} \, cm^2/s$. The results in **Figure 3b** show that the experimental result reported[33a], 74% lithiation for monodisperse particles with a radius of 250 nm at 5C, can be reproduced by the present phase-separating phase field model, resulting in 72% lithiation, using measured diffusion coefficients.

### 3.2. Validation of the model, impact of electrode thickness and porosity.

In order to further validate our model, multiphase porous electrode calculations are compared to several prepared electrodes, summarized in **Table 2**. In **battery 1** the LTO electrode is relatively thin (20 μm) and highly porous (70%). As a consequence the Li-ion diffusion and the phase transition behavior inside the LTO can be expected to dominate the performance, which allows further validation of the present LTO phase field description. To validate the charge transport description in the model through the porous electrodes, two additional electrodes were prepared where thickness (**battery 2**) and porosity (**battery 3**) were varied (**Table 2**). All LTO electrodes were tested in half-cells vs Li metal, as described in the methods section, and have a relatively large carbon content (15%) to be entirely sure that the electronic conductivity does not contribute significantly to the internal resistance. The particle size distribution in the porous phase field model was determined based on SEM images of the LTO material, shown in **Supporting Information A**. All the relevant parameters for the phase field model of **battery 1** are shown in **Table 1** while the additional morphology aspects (porosity and thickness) are provided in **Table 2**. The phase field model predictions for **batteries 2** and **3** were performed before actual electrodes were prepared and for the simulations of the three batteries no parameters were fitted.

**Table 2:** Electrode characteristics of **battery 1**, **2** and **3**.

| Aspect | Battery 1 | Battery 2 | Battery 3 |
| --- | --- | --- | --- |
| Aim | Switch off electronic and ionic transport | Test Li-ion transport behaviour due to electrode thickness | Test Li-ion transport behaviour due to electrode porosity |
| Experimental Thickness ($L_c$) | 20 μm | 245.5 μm | 225.5 μm |
| Experimental Porosity ($\varepsilon$) | 70 % | 67% | 52% |
| Simulated Thickness ($L_c$) | 20 μm | 250 μm | 250 μm |
| Simulated Porosity ($\varepsilon$) | 70 % | 70% | 50% |

Comparison between the experimental and the simulated rate capability for **battery 1** in **Figure 4a** shows the characteristic declining capacity "staircase" upon increasing C-rate. The experimentally obtained and simulated capacities are in excellent agreement with each other over almost 3 decades in discharge rate, showing just 2% difference in the capacities.

However, the porous phase field model has a mismatch with the overall shape of experimental voltage curves shown in **Figure 4b**. Clearly, our model has overly high reaction resistance at low concentration since the initial voltage drop, prior to any phase separation or diffusion limitation, is much larger than the experimental results. After the initial drop, the experiments show steeper voltage decline than the phase-field simulations. Below we will address several reasons that might be responsible for such behaviour and to what extent this discrepancy may signify incomplete physical understanding.

One reason may be the breakdown of the standard assumption of spherical symmetry, which promotes shrinking-core phase separation within each active particle. In the case of LFP, the canonical two-phase material, it is well established that the initial shrinking core model[56] cannot capture various surface phenomena predicted by phase-field models that relaxed this assumption[18-21], such as the electro-autocatalytic control of phase separation by surface reaction kinetics[23, 30], which was later verified experimentally[24] with definitive proof by *in operando* imaging of nanoparticles[29]. It is also important to note that a 1D phase field model of LFP nanoparticles[57] similar to our LTO model, which allowed for surface nucleation and departures from shrinking core morphologies, such as radial spinodal decomposition and annual stripes, also produced flat voltage profiles resembling ours for LTO and failed to capture the observed rate-dependent phase behaviour in LFP.

Let us briefly discuss how departures from spherical symmetry that allow for electro-autocatalytic control of phase separation[30] may help explain varying voltage profiles in LTO. As discussed earlier, DFT results indicate that the phase interfaces are stabilized by mixed 8a/16c Li site occupation that resemble solid solution conditions on top of the nanoscale two phase separation, explaining the fast Li-ion mobility and phase-interface stabilization in LTO.[16] This picture was recently confirmed by a thorough experimental (*in situ* XANES) and computational (*ab initio*) study pointing towards a reaction process that involves mixed quasi-solid solution/macroscopic two-phase transformations.[58] As a result, deviations from a shrinking core phase separation are expected in reality. The material may not display one uniform phase front (core-shell) but several, thus exposing Li-poor phases close to the particle surface. It is nontrivial and beyond the scope of this initial work to capture driven phase separation in 2D or 3D, although that is a key direction for future work. Since the penalty to form phase interfaces is extremely small, nanoscale alternating Li-rich and Li-poor phases would be in principle possible. For example, a chessboard-like, nano-domain lithiation mechanism has been predicted for graphite[31a]. Further, if phase separation can occur in the tangential direction over all the surface (contrary to shrinking core), then it is strongly influenced by surface reactions. Whenever the reaction is auto-inhibitory during insertion as predicted by our theory, then phase separation is suppressed during insertion, and the overpotential grows over time, leading to voltage curves that drop off more like the experimental data. We have growing unpublished evidence that this effect is important in multiple battery materials, not only suppressing, but in some cases enhancing concentration instabilities (i.e. making a solid solution system appear to phase separate at high rates into non-equilibrium states of stationary reaction rate), as predicted by the general theory[18a]. Furthermore, a simpler reason for breaking the spherical symmetry might be the non-perfect character of the real crystals, meaning that the grain boundaries might affect the phase boundaries, leading to local growing/shrinking core shell structures. In this case larger particles may not display one uniform phase front (core shell) but several, having more Li-poor regions near the surface, and thus lower overpotentials.

Another reason for the different shape of the voltage curves could be the breakdown of standard assumption of Butler-Volmer reaction kinetics, as was recently implicated in the similar case of LFP[34]. A different, physics-based alternative to the empirical Butler-Volmer equation is provided by quantum mechanical theories of electron transfer[59], such as the Marcus-Hush-Chidsey model (MHC)[60] which predicts an electron-transfer reaction-limited current that cannot be exceeded even at high over-potential[61] (opposite to BV theory, which predicts faster kinetics at higher overpotential). A generalized theory of mixed ion-electron transfer in solids[62], combining MHC kinetics with nonequilibrium thermodynamics[18], predict an approximately linear decrease in exchange current vs concentration, $I_o \sim 1-c$, for mixed ion-electron transfer into a solid intercalation material, consistent with experimental measurement of $I_o(c)$ in LFP[30], which can explain the linearly decreasing capacity at low rates. A preliminary result from implementing the generalized MHC theory in our LTO phase-field model is presented in **Supporting Information B**, which indeed shows a steeper decline in the voltage curve, closer to the experimental data.

Another possible important extension of our model would be to take into account many particle effects within the continuum volumes. In this way the spatial nonuniformity of electrochemical resistances can be captured, causing a distribution of voltages. Particles within one volume (same electrode depth) have a unique (higher or lower) connection to the average electrode potential, based on their unique aspects (tortuosity, crystalline orientation, size, particle-particle contact)[28]. For example, the inclusion of a contact resistance in phase field modeling resulted in good voltage fitting in the case of graphite.[28] In addition, the more sloping nature of the experimental voltage curves can be attributed to a voltage distribution due to the size distribution of the LTO powder not taken into account by the current phase field model. For small particle sizes the relatively large surface area makes that the surface energy increasingly changes the potential. As a consequence a size distribution leads to a distribution of insertion potentials[63]. Finally, another approach would be to use more complex empirical functions for the concentration-dependent solid diffusivity. Such functions have been shown to fit the LFP voltage profiles quite well[56], and might lead to a more accurate fitting of the LTO voltage profile as well. This is however outside the scope of this work, which aims at the development of the first LTO phase field model without fitting parameters, and very practically and accurately captures the lithiation properties of LTO based on its fundamental thermodynamic properties.

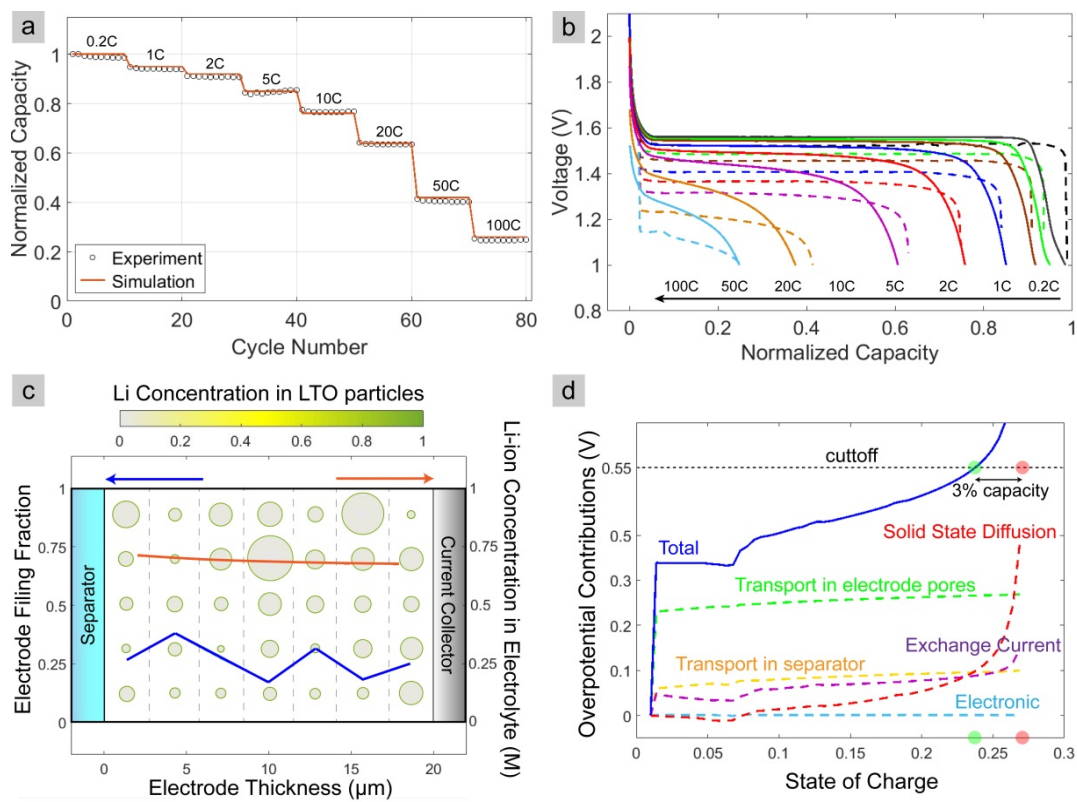

**Figure 4: a)** Comparison between the experimental (scatter) and the simulated (line) capacities for **battery 1**. **b)** Experimental (solid lines) and simulated (dashed lines) voltage curves at different C-rates. **c)** Li-ion end concentration in the particles, the average filling fractions in the solid phase (blue line - left axis) and Li-ion concentration in the electrolyte (orange line - right axis) plotted as a function of electrode depth at the end of a 100C simulation for the **battery 1** half-cell. **d)** Overpotential contributions to the total overpotential as a function of the state of charge for 100C discharge of **battery 1**.

To investigate what limits the capacity at large C-rates, the Li-ion distribution through the electrodes and in the LTO particles, as well as the origin of the overpotential is analyzed in detail. In **Figure 4c** the Li concentration profile in the solid phase and in the electrolyte with respect to electrode depth is presented for 100C at the fully discharged state for **battery 1**. The simulations indicate that at 100C, all particles are lithiated concurrently throughout the electrode. The concentration profile in the solid phase is dominated by the distribution of particle sizes along the depth of the electrode, with smaller particles at a larger state of discharge. The average Li-ion concentration in the particles is nearly constant throughout the electrode depth. This qualitatively indicates that for this thin (20 $\mu$m), porous (70%) electrode Li-ion transport in the solid LTO particles limits the Li-ion transport and the resulting capacity. The various contributions to the overpotential for **battery 1** are quantified as a function of state of discharge in **Figure 4d**. Although **Figure 4c** shows that the electrolyte is only partially depleted, the Li-ion transport through the electrolyte in the pores of the electrode already contributes significantly to the overpotential. At these large rates the Li-ion concentration decreases, gradually increasing the overpotential. This gradually decreases the voltage until the cutoff potential is reached, thus the capacity is mainly limited by the Li-ion transport through the electrolyte. In contrast, at lower rates a sharper decrease in discharge voltage is observed. The origin of this is illustrated in **Figure 4d,** in which the overpotential contributions are displayed for 100C rate beyond the 1 V cutoff. Just beyond the voltage cutoff the overpotential due to Li-ion diffusion through the LTO particles sharply increases. At lower rates this regime is entered before the 1 V cutoff is reached because the overpotential due to Li-ion transport though the electrolyte is smaller. Therefore the shape of the voltage profile towards the voltage cutoff can be directly used to identify which transport phenomena limits the total capacity, which may serve as guide to improve the performance. Specifically for **battery 1**, **Figure 4c** indicates that improving the charge transport though the electrolyte, for instance by a larger Li-ion concentration or larger porosity, will only lead to a gain of about 3% in capacity. At that stage the Li-ion diffusion through the LTO particles sets in, indicating that particle size limits the capacity.

According to the analysis presented in **Figure 4c**, electronic limitations do not play a significant role for the electrode in **battery 1** for which a conductivity of 10 S/m was set. To evaluate the impact of the electronic conductivity simulations were performed with lower electronic conductivities, shown in **Supporting Information C**, **Figure S5.** Reducing the electronic conductivity to 0.3 S/m shows a minimal effect on the capacity (~2%), despite the high current (100C).

The next step is the validation of the Li-ion transport behaviour of the porous phase field model by varying the electrode thickness and porosity in two subsequent steps. As summarized in **Table 2,** compared to **battery 1** the porosity of **battery 2** is kept constant, but the electrode thickness is increased from 20 μm to 250 μm. For **battery 3** the electrode thickness is kept the same as for battery 2 and the porosity is decreased. For **battery 2** the simulated and the experimentally obtained voltage profiles are shown in **Figure 5a**. The model predicts that above 5C the voltage drops quickly towards the cutoff voltage, leading to a sharp decrease in capacity at 10C, in good agreement with the experimental result. The predicted capacities for different discharge rates are in excellent agreement with the experiments (within 4%). In **Figure 5b** the simulations at 10C display an interesting distribution of Li-ions over the LTO particles as a function of depth, and a clear gradient in the Li-ion concentration in the electrolyte. In contrast to the thin electrode of **battery 1** in **Figure 4c**, this indicates that in these much thicker electrodes Li-ion transport through the electrolyte contributes more to the overpotential than Li-ion diffusion trough the LTO particles, a direct consequence of the larger diffusion distance through the electrolyte. As a consequence, the discharge capacity is dominated by the Li-ion transport down to much lower rates. This is qualitatively observed by the gradual decrease in voltage during discharge (indicating that the Li-ion electrolyte transport dominates the overpotential) as compared to the sharp decrease in voltage (indicating Li-ion diffusion in the LTO particles dominates the overpotential). By quantifying the overpotential contributions coming from the various kinetic mechanisms to the total overpotential in **Figure 6a** we can directly observe the dominant character of Li-ion transport in the electrode behaviour at 10C discharge rate. The contribution of electronic transport seems to play a more significant role compared to **battery 1**. This makes sense if we consider that the Ohmic resistance runs over the larger length of **battery 2**. Evidence that for this thick electrode (**battery 2**) the electronic wiring is an issue is also visible in **Figure 5b,** where a reversed Li concentration gradient can be seen in the solid phase near the current collector side. To investigate the impact of the electronic conductivity of **battery 2**, the electronic conductivity was varied at 10C discharge, the results of which are shown in **Figure S6** in the **Supporting Information C**. Starting at 10 S/m, which was used in the simulations shown in **Figure 5a**, reducing the electronic conductivity steeply reduces the capacity and increases the electronically induced overpotential, while larger electronic conductivity values (100 S/m) lead to better capacities and less overpotential.

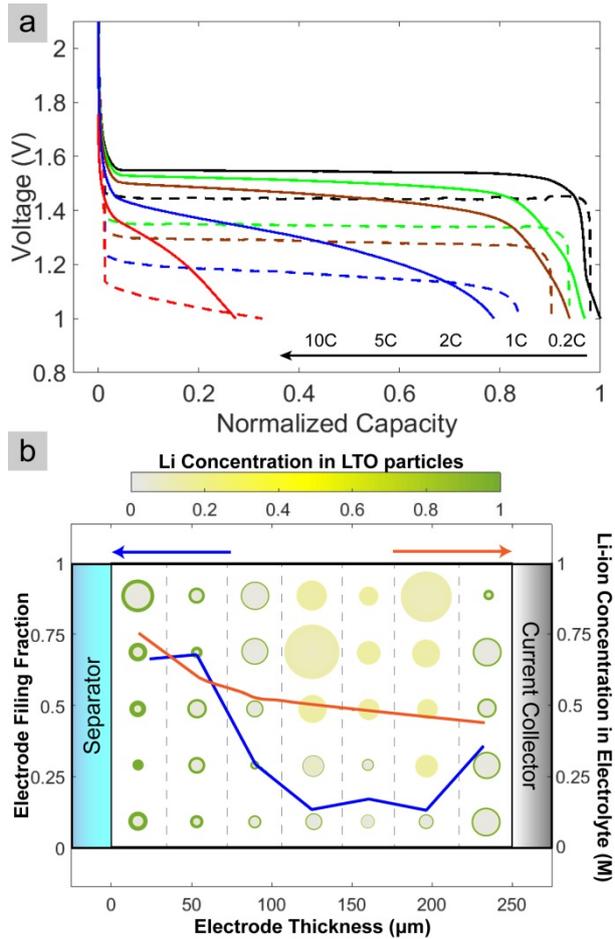

**Figure 5: a)** Simulated (dashed line) and experimental (solid line) voltage profiles of **battery 2 b)** Li-ion end concentration in the particles, the average filling fractions in the solid phase (blue line - left axis) and Li-ion concentration in the electrolyte (orange line - right axis) plotted as a function of electrode depth at the end of a 10C simulation for the **battery 2** half-cell.

**Battery 3** has a similar thickness compared to **battery 2**, but a lower porosity, which is beneficial for the energy density of a full battery but compromises Li-ion transport through the electrolyte in the pores of the electrode. We initially performed simulations in order to investigate the effect of porosity on a 250 $\mu m$ thick electrode. The impact of the porosity on the capacity, at a cutoff voltage of 1 V, is summarized in **Figure 6b**. Also for **battery 3** the agreement with the experimentally obtained capacities are excellent. As expected, decreasing the porosity dramatically reduces the capacity at high discharge rates. Lowering the porosity induces a larger overpotential due to the Li-ion transport through the electrolyte, enhancing the effect seen and discussed in **Figure 6a**. In transport limited systems such as **battery 2** and **3**, total capacity is more likely to be determined by the balance between the overpotential build up and the cutoff voltage (as shown for example in **Figure 6a**). In this case, the voltage mismatch, observed in **Figures 4**b and **5a**, might be a cause of concern, but our simulations show that this is not the case. Despite the initial overestimation of the overpotential, accurate description of the rate limiting mechanisms and total capacities is achieved even at extreme electrode morphologies (250 $\mu m$ thickness).

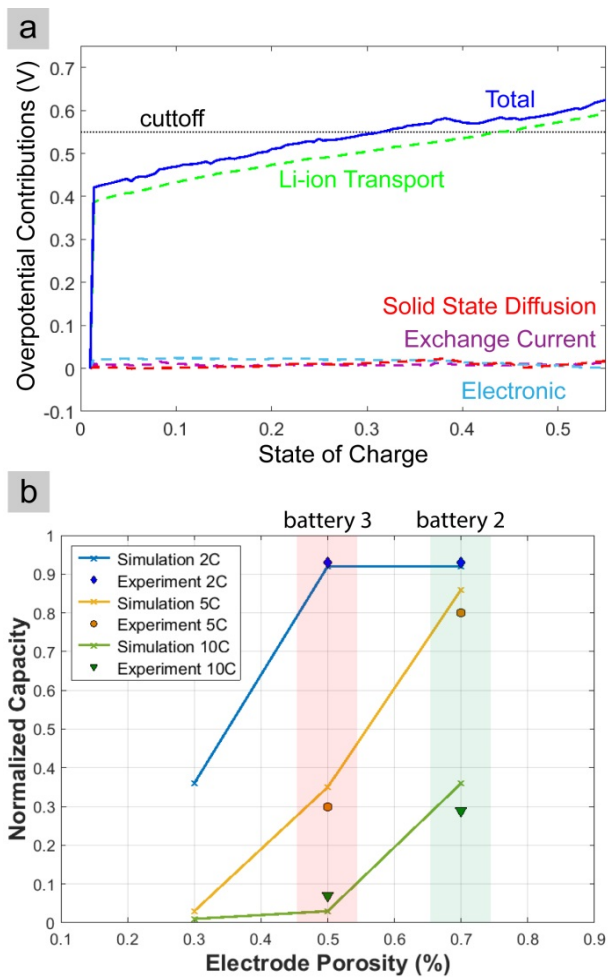

**Figure 6 a)** Overpotential contributions to the total overpotential as a function of the state of charge during a 10C discharge of **battery 2**. **b)** Simulated vs. experimental capacity obtained as a function of porosity for a 250 μm thick electrode at various rates.

Moreover, in **Supporting Information D**, the LTO model is compared with a literature study in which the carbon content and electrode thickness are systematically varied. In addition, the particle size utilized is different from the one used for **batteries 1,2**, and **3**. In this way, the adaptability of the model towards different electrode morphologies is demonstrated. The model is able to explain the equivalent electrochemical behaviour[46, 47] observed when cycling, up to 10C, for LTO electrodes with and without carbon.

Evaluating the results presented in this section, we conclude that a reasonably accurate electrode description may be achieved with the well-established modeling assumptions (BV and shrinking core) and the practicality of a 1D model, despite the physical limitations discussed above. Because of the differences in the voltage curve, a different voltage cutoff would produce different results for the simulated vs. experimental results, in the case of transport/transfer limited system (e.g. flatter simulated line would lose much more capacity than the curved experimental line for a cutoff at 1.2V). Cutoff at 1V, however, is the only relevant cutoff voltage encountered in literature and commercial applications. Despite the potential mismatch, capacity behavior is reproduced for extreme cases of currents and electrode morphologies. Even when extreme transport and transfer electrode characteristics are limiting the electrode (battery 2 and 3), capacities are well reproduced. An exact match of the voltage curve, however, would be the next step in the LTO phase field model, where including more realistic models of mixed ion-electron transfer, a better morphological description and/or multidimensional modeling might be the key elements for an improved material description.

### 3.3. Electrode Optimization

The good agreement with experiments shown in the previous section motivate the use of the developed porous phase field model to predict and explain the macroscopic electrochemical performance of LTO electrodes. In the next section we use the predictive power of the model to systematically evaluate the capacity as a function of particle size and particle size distribution, as well as the electrode thickness and porosity for different C-rates, providing guidelines for optimal LTO electrode design. **Figure 7a** shows the maximum capacity that can be obtained in single particles over a wide range of particle sizes and C-rates, hence for the conditions in which Li-ion transport through the electrolyte and electronic conductivity do not play a role. For example, when aiming at a maximum energy density battery operating at 2C, **Figure 7a** indicates that particles with a radius of 250 nm would give more than 90% of the theoretical capacity. **Figure 7a** shows that LTO nanoparticles (radius < 40 nm) can be lithiated completely, even at rates of 100C, consistent with experiments.[2d, 41e, 64] When Li-ion diffusion through the LTO is at the nanoscale (5 < R < 50 nm),[2d, 41e, 64] and the contribution of the Li-ion transport through the electrolyte and electronic conductivity to the overpotential are minimized (typically by high porosities, high surface areas, thin electrodes and the use of conductive coatings), LTO can indeed reach a high fraction of the theoretical capacity at 200C[105]. In **Figure S10a, Supporting Information E** a capacity map, similar to **Figure 7a**, is provided with the simulation results for small nano-crystallites and extremely high discharge rates. In addition, we report various half-cell scenarios in **Figure S10b** aiming at high capacity retention at ultra-high discharge rates.

The limitations presented in **Figure 7a** most likely hold true for diffusion within particles consisting of single grains. This is quite important as particles consisting of multiple grains would behave differently. Generally, it has been shown that the introduction of more grain boundaries and thus phase interfaces enhances the Li-ion and electronic conductivity, ultimately boosting the electrochemical performance.[16, 41d, 65]. There are many studies that aim at creating larger secondary particles consisting of smaller primary particles, showing superior electrochemical behaviour.[2d] For example, Wang et. al.[41d] managed to synthesize monodisperse secondary particle spheres with a radius of 220 nm that consist of nano-sized primary particles which perform extremely well (68% capacity at 30C). This capacity is doubled compared to the limitations indicated in **Figure 7a** and is attributed to the large number of transforming grains that will effectively create additional interfaces between the end member phases, thus enhancing diffusion.[15, 16, 39a, 41d]

It has also been argued that the macroscopic two-phase separation falls apart in nano-domains under equilibrium conditions[15], or even the possibility for a percolation mechanism with multiple phase interfaces during the lithiation process[65, 66]. As thoroughly discussed in the previous section our 1D model does not indicate the formation of multiple phase-interface fronts although a better description could be achieved by adding more dimensions.

As the LTO particle size can be decisive for the rate dependent performance, also the size distribution is expected to have a significant impact on the rate dependent capacity[2d, 41e, 42]. An indication of the capacity decrease as a result of the presence of larger particles can be obtained from **Figure 7a**. In the case of lognormally distributed particles large particles make up a large part of the electrode solid volume, **Figure S11** in **Supporting Information F** compares the capacity obtained from a log-normal particle distribution around a mean particle radius of 100 and 50 nm with different standard deviations. This demonstrates how a wider particle size distribution lowers the capacity, underlining the importance of monodisperse electrode particles to maximize the capacity of Li-ion battery electrodes. A sharper drop in capacity is predicted for particle size distributions with a smaller average size because of the relatively larger impact of the largest particles in such a distribution. In conclusion, the simulations show that avoiding large particle aggregation is of primary importance to achieve improved capacity retention at high discharge rates.

The porous electrode simulations, shown in **Figures 7b, c, d, e** and **f** illustrate how electrode thickness, particle size, porosity, tortuosity and the separator influence the capacity as a function of C-rate. For these simulations a realistic active particle fraction of 80% active material, 10% PVDF binder and 10% carbon black is used, resulting in an active material volume loading of 69%. The Celgard 2500[67] separator was implemented because it is often employed in both experimental and modeling studies. In **Figure 7b**, simulations were performed for monodisperse particles with a radius of 100 nm and the porosity of the electrode was fixed at 26%, a typical commercial porosity for electrodes, and the electrode thickness as well as the C-rate was varied. The declining capacity with increasing electrode thickness is the result of Li-ion transport limitations through the electrolyte in the electrode pores. For example, a single particle with a radius of 100 nm is expected to give more than 80% capacity at 20C (**Figure 7a**). However, for a porous electrode this is only achieved for extremely thin electrodes because with increasing electrode thickness the Li-ion transport through the electrolyte will increasingly dominate the overpotential. This is also demonstrated in **Figure 7c,** where the electrode thickness is fixed at 100 $\mu$m and the particles size is systematically varied. When comparing with the single particle simulations in **Figure 7a** this shows that particle size is almost irrelevant for thick electrodes. For example at 5C discharge rate the capacity dependence is dominated by transport, the capacity is almost constant and equal to the one determined by electrode thickness as depicted in **Figure 7b,** and nanosizing (reducing the radius to 50nm) has no effect on the capacity.

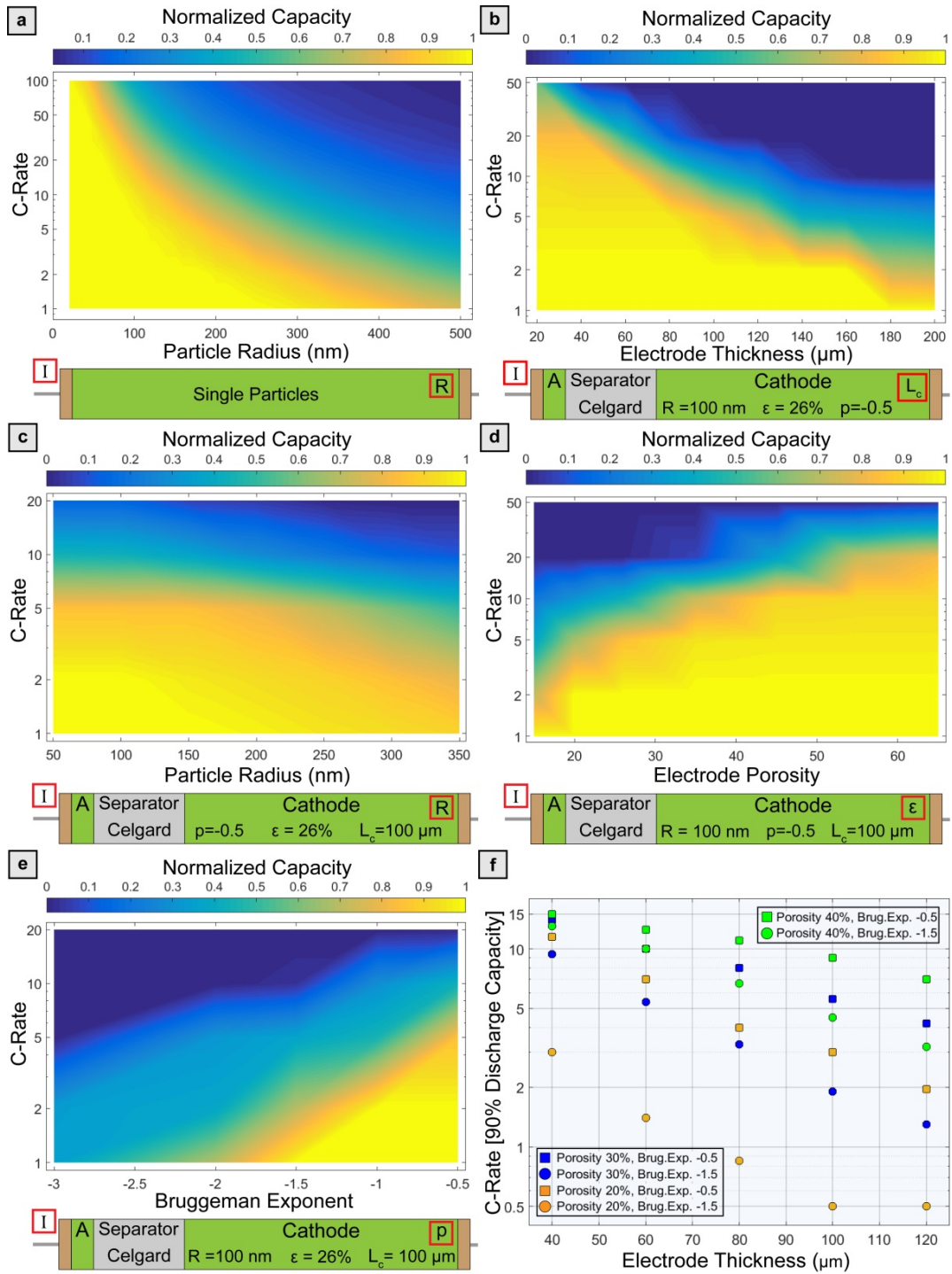

**Figure 7: a)** Capacities obtained from LTO single particle simulations over a wide range of particle sizes and C-rates. **b) c) d) e)** Capacities obtained from LTO porous electrode simulations. The electrode properties are shown in the schematic electrode representation below each figure. The parameters in the red squares are the ones being varied. **f)** Maximum C-rate delivering 90% of the nominal discharge capacity as a function of electrode thickness for different combinations of porosity and Bruggeman exponent. The particle radius is 100 nm.

In **Figures 7d** and **e** the electrode thickness is kept constant at 100 μm and the porosity and tortuosity were varied. Tortuosity, as introduced in the **2.1.2 methods** section, appears to have a really strong influence on the capacity. The ideal Bruggeman exponent is -0.5, which refers to monodisperse spherical particles, emphasizing the importance of a synthesis process able to achieve spherical morphology and monodispersity.

This is better demonstrated in **Figure 7f,** in which parameters that seem to have the biggest influence on capacity are combined into a 5D plot. It shows the maximum C-rate delivering 90% of the nominal discharge capacity as a function of electrode thickness for different combinations of porosity and Bruggeman exponent. The particle radius was 100nm. As an example, we observe that the less torturous 20% porous electrode outperforms the much more torturous 30% porous one, showing the potential gain in energy density.

In **Figure S12a**, **Supporting information G** a similar heat-map to **Figure 7b**, where all parameters are kept the same except from a higher porosity of 40% for those aiming at high rate performances, is presented. Because of the additive character of the overpotentials, the choice of the separator will most likely affect battery performance in transport limited systems. We demonstrate this in **Figure S12b**, **Supporting Information G** were for comparison a heat map is created implementing the Whatman glass fiber (GF) 260 separator[52], which is also a common choice in literature and was the separator used for **batteries 1, 2** and **3**. Results indicate that the extremely small thickness (25$\mu$m) and medium porosity (55%) of the Celgard 2500 separator outperforms the large thickness (200$\mu$m) and high porosity (87%) of the Whatman GF, especially at high C-rates. This explains why in literature we mostly encounter Celgard 2500 separators in papers aiming at high rate performance. Since the main characteristics of a separator are described in terms of porosity and thickness this result can be generalized and can be compared with other separators.

Based on the LTO phase field model simulations for **batteries 1 to 3** a Ragone plot was calculated (**Figure 8**). It is based on the present phase field model for a LTO anode combined with a high voltage spinel $LiNi_{0.5}Mn_{1.5}O_4$ (LMNO) cathode, assuming that the LTO anode limits the performance. The weight of the full cell was determined by matching the LMNO and LTO capacities. We also assumed equivalent potential losses for the two electrodes producing a constant voltage difference of 3.15 V (4.7 -1.55). [68]. For scaling the batteries towards realistic practical energy and power densities, scaling factors based on the characteristics of each electrode were used with the lowest value (17%)[69] for the highly porous and thin electrode (**battery 1**) and the highest (44%)[69] for the thick and less porous one (**battery 3**). The results in **Figure 8** are comparable with that reported for a $Li_{4+3x}Ti_5O_{12}/Li_{y+0.16}Mn_{1.84}O_4$ cell. [33b]

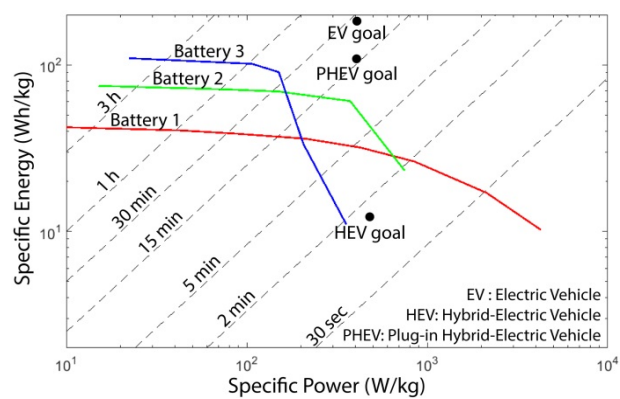

**Figure 8:** Ragone plot of LMNO – LTO batteries, the dashed lines indicates the particular discharge times. Batteries 1, 2 and 3 refer to full cells (LMNO – LTO) where the characteristics of the LTO electrode are the same with the ones presented for **battery 1, 2** and **3** respectively. The practical energy and power density was determined by scaling **battery 1** with 17%, **battery 2** with 30% and **battery 3** with 44%. The EV, PHEV and HEV goals are determined by the U.S. Department of Energy commercialization goals[33b].

### 3.4. Active Particle Fraction in LTO electrodes

An important aspect of the performance of complete electrodes is the active particle fraction. When the active particle fraction is larger, the total current is distributed over a large fraction of the particles. However, when the active particle fraction is small, the current is more localized, which could indicate hot-spots that may be detrimental for the cycle life of batteries. Li et. al. investigated both computationally and experimentally the active particle fraction in LFP electrodes, reporting a strong dependence on cycle rate and direction.[27] It was observed that in case the current exceeds a certain rate (~9C) the LFP electrode attains concurrent Li-ion insertion. In this way they bridged contradictory reports suggesting particle-by-particle[53, 70] and concurrent[71] insertion behaviour in LFP electrodes.[27] For any phase-separating active material, such a range of rate-dependent phase transformation pathways is a general prediction of multiphase porous electrode theory[18b, d, 19].

With the LTO phase field model we investigated the active particle fraction of the **battery 1** LTO electrode as a function of C-rate and state of charge. The criterion applied for defining a particle as actively transforming is the formation of a propagating phase interface. Our findings indicate that the active particle fraction shows a strong dependence with increasing rate, very similar to what was predicted for LFP electrodes.[27] In addition, **Figures 9b** to **e** demonstrate that the active particle fraction strongly depends on the state of charge. This effect was quantified by plotting the active particle fraction in terms of the number, surface and volume of particles that are active vs. the state of charge during discharge in **Figure 9b** to **e**. In **Figure 9a** the active particle fraction (average value for each C-rate) and current density as a function of rate is presented. Simulations demonstrate that at the initial stages of lithiation all particles lithiate in a solid solution manner (note that this does not qualify as active according to our definition and thus shows up as zero activity in **Figure 9b** to **e**).

The prediction that the LTO particles remain in a solid solution lithiation state, above the 3% expected based on the regular solution model, is the consequence of an absence of a thermodynamic driving force to exit the metastable region, as discussed in **section 3.1**. Even though all particles begin the lithiation process with a solid solution mechanism, the number of particles that will continue and begin the two-phase transformation towards the $Li_7Ti_5O_{12}$ phase strongly depends on the C-rate. This effect is also predicted for LFP electrodes and is explained by the presence of a transformation barrier in phase separation electrodes. This barrier is defined by the potential difference between the local maxima and the middle point of the diffusional chemical potential of Li in the phase separating material particles.[27] When the electrode ensemble potential is below the activation barrier, particles are lithiated up to the spinodal point (the end-point of the metastable region), but when the barrier is crossed the particles activate and lithiate further.[27] Transformation at low rates occurs in a strict sequential order, where small particles are lithiated first (particle-by-particle lithiation). This is evident in **Figures 9b** to **d,** where at the initial stages of discharges the number of active particles and active particle surface that are active is higher than the active volume fraction, while this situation is reversed in the latter stages of discharge. In **Figure 9a** a sharp rise in the active particle fraction is observed for currents around 9C, as the material gradually shifts form a particle-by-particle to a concurrent lithiation mechanism. Small particles still have a small edge in the lithiation order, but as the overpotential driving force[27] is increasing with increasing C-rate, more and more particles are lithiated concurrently and at 20C rate all of the particles are actively transforming. In retrospect, in terms of active particle fractions LTO behaves similar to LFP, which suggests that these results are general for phase separating electrodes, as predicted theoretically[18b, d, 27], although more analysis is required to support this conclusion. For LTO, a particle-by-particle reaction mechanisms was proposed by Kitta et. al. at low rates (0.3C) via electron energy loss spectroscopy[54] which is in agreement with our simulations of the active particle fraction at 0.2C rate in **Figure 9b**. We present (**Supporting Information**) the electrode lithiation behavior in a series of movies for various rates of both **battery 1** and **2**, demonstrating the lithiation mechanism, along with the evolution of the Li-concentration in the electrode and electrolyte during discharge.

The prediction that at low currents only a very small fraction of the particles carry the current may lead to a maximum in the surface current density at a certain C-rate, which can lead to local hot spots that enhance electrolyte degradation locally. By taking the average active surface fraction of each C-rate we constructed a plot quantifying this behaviour for **battery 1** in **Figure 9a**. We observe that such a maximum is not present, implicating that the increasing C-rate is the dominating factor, even though the active particle fraction scales proportionately with the C-rate.

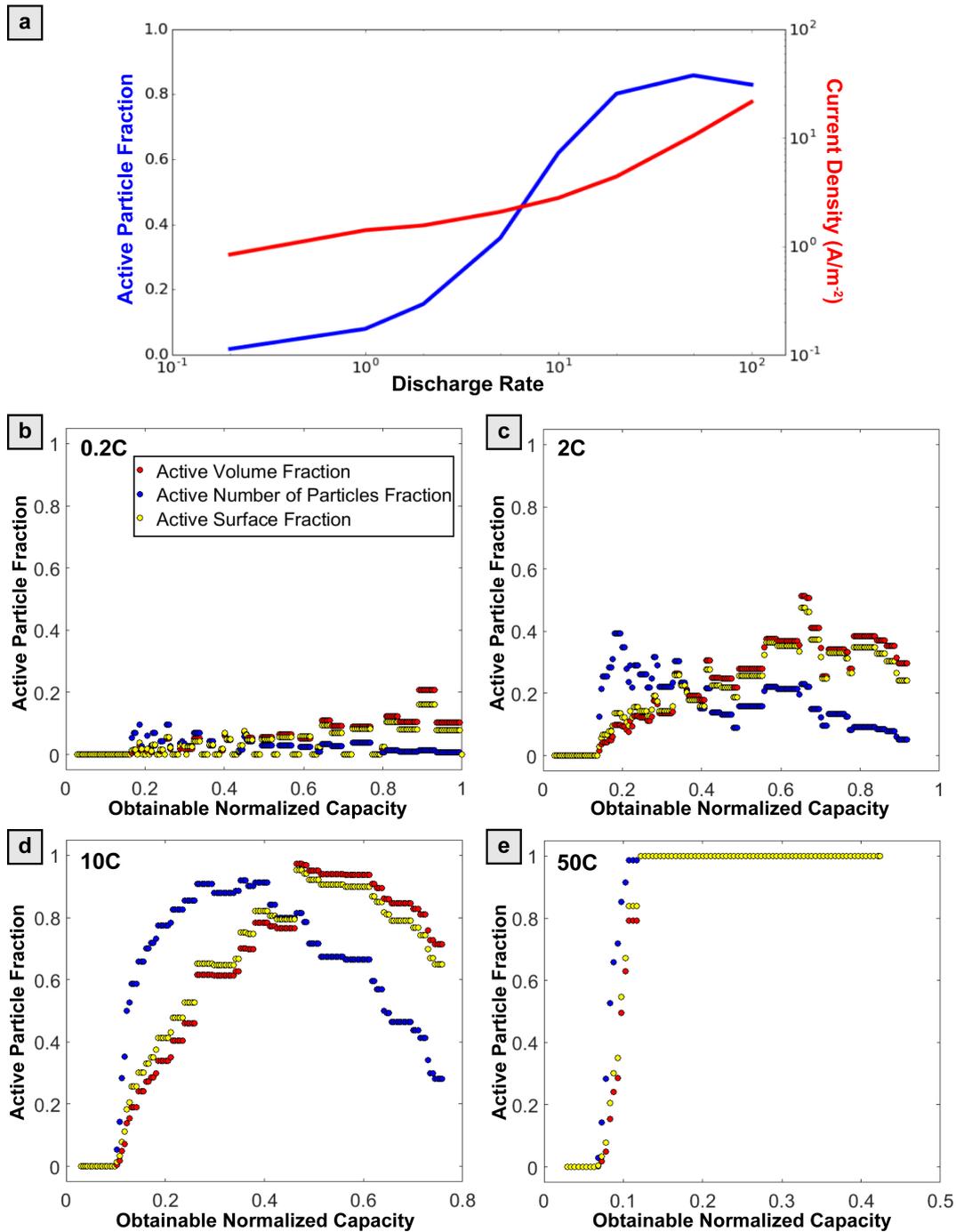

**Figure 9: a)** Active particle fraction (based on the active surface) and current density of **battery 1** as a function of rate. Active particle fraction (based on active volume, surface and number of particles) vs. obtainable normalized capacity of **battery 1** for various rates **b)** 0.2C **c)** 2C **d)** 10C **e)** 50C.

Finally, we examine if the prediction of a concurrent solid solution reaction anticipated for the initial stages of lithiation has a physical dimension. Song et. al.[46] and Kim et. al.[66] performed electrochemical tests in carbon-free electrodes. Both groups observed a color change from white to black in the LTO electrode, indicating a sharp change of the conductive properties, before the start of the characteristic voltage plateau. Thus it was suggested that the phase equilibrium does not take place from fully insulating $Li_4Ti_5O_{12}$ particles as believed, but from an already conducting $Li_{4+d}Ti_5O_{12}$ phase.[46] This means that LTO becomes conductive before the two-phase transition begins. In addition, Kim et. al.[66], visualized the phase propagation through optical microscopy and µ-XAS. An initial fast lithiation wave was observed that occurs throughout the electrode at the very beginning of the lithiation process, making it conductive, followed by a secondary wave of bulk lithiation.[66] The second wave of bulk lithiation follows a particle-by-particle lithiation mechanism, it begins at the regions in closest contact with the current collector as the reaction remains somewhat sluggish.[66] The present simulations are in agreement with the above descriptions. The concurrent character of the solid solution insertion observed in the phase field model at the initial stages of lithiation could explain the experimentally observed[66] initial lithiation wave that occurs throughout the electrode, making the material conductive. This initial wave is followed by a secondary lithiation wave in the bulk (phase separation in the particles), the sequence and concentration gradient of which depends on the particle size, wiring and C-rate. Electrode simulations with zero-carbon content are presented in **Supporting Information D**, where the proposed mechanism[47, 66] of the secondary bulk lithiation beginning close to the current collector which propagates towards the electrolyte side is discussed.

## 4. Conclusion

The developed LTO phase field model can accurately reproduce and predict electrochemical behavior. The model integrates DFT data in an effort to include a complete thermodynamic description that leads to phase separating particles. It captures the fundamental physics of LTO electrodes, while explaining and predicting battery performance and the various rate-limiting factors. The study covers a wide rate of C-rates, particle sizes, electrode morphologies and different separators. In addition, guidelines for optimized electrode performance are given. The importance of monodispersity is emphasized, along with the importance of the balance between transport and electronic effects. Finally, insight is provided regarding the LTO lithiation mechanism by a thorough examination of the active particle fraction. A particle-by-particle lithiation mechanism is confirmed for low C-rates, in agreement with experiments, and above approximately 9C a concurrent lithiation mechanism is predicted, a behavior which is similar to the one observed for LFP electrodes.

**Supporting Information**

Supporting Information is available from the Wiley Online Library or from the author.

**Acknowledgements**

The research leading to these results has received funding from the European Research Council under the European Union's Seventh Framework Programme (FP/2007- 2013)/ERC Grant Agreement No. [307161] of M.W. Financial support from the Advanced Dutch Energy Materials (ADEM) program of the Dutch Ministry of Economic Affairs, Agriculture and Innovation is gratefully acknowledged. R.B.S. and M.Z.B. acknowledge partial support from Samsung-MIT Alliance and the D3BATT project of the Toyota Research Institute.

TABLE OF CONTENT (TOC)

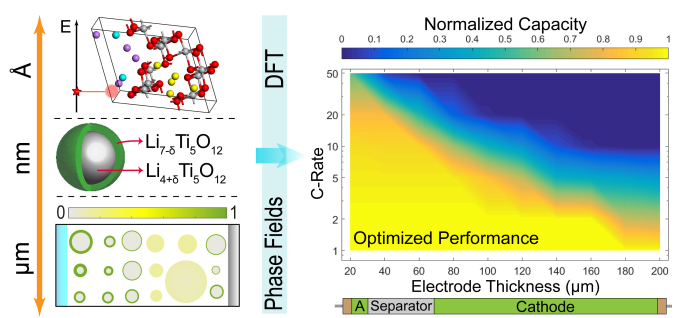